# 1. Introduction

A plasma torch is a non-equilibrium plasma source operating at atmospheric pressure, generating a non-thermal flowing post-discharge. It is supplied with a carrier gas (usually helium or argon) that can be mixed with a reactive gas (usually oxygen or nitrogen). In recent years, plasma torches have been actively investigated as an effective means for the surfaces treatment such as etching, surface modification, sterilization, … [Choi-2005], [Shenton-2001], [Fang-2003], [Gonzalez-2010]. They have also become an alternative approach to the atmospheric plasma jets in biomedical applications such as the treatment of living tissues and cells [Yonson-2006], [Landsberg-2010], [Nosenko-2009], [Kuo-2009], [Kong-2009]. Understanding the properties of a treated surface (surface energy, composition, …) requires a complete understanding of the post-discharge/surface interactions and a background characterization of the mechanisms ruling the flowing post-discharge.

Since a decade, atmospheric He-O$_2$ plasmas have been extensively studied but only a few papers are focused on the experimental characterization of their flowing post-discharges [Gudmundsson-2004], [Jeong-2000], [Cardoso-2006], [Chan-2011], [Chiu-2010]. Simulations predicting their behavior are also quite complicated because of the outstanding number of reactions occurring at atmospheric pressure, because of the interaction of the species (radicals, ions, metastables, dimmers, …) with the atmospheric air and finally because of turbulences phenomena [Goree-2006].

In this paper, we have characterized a "curtain post-discharge" generated by a scanning RF (27.12 MHz) plasma torch, fuelled with helium and oxygen. The design of the plasma source is original in that the gap between the two flat electrodes is only 1 mm. Moreover its nozzle's section is linear (20 mm by 0.8 mm), thus giving to the post-discharge a curtain geometry. We will first examine the influence of the He-O$_2$ flow rates on the post-discharge behavior. Despite an RF running mode of the plasma, we will also stress the existence of a DC current (1-20 µA) measured within the post-discharge. The mechanisms governing this DC current are investigated through electrical measurements, mass spectrometry and optical emission spectroscopy. We have also demonstrated how the current measured in the post-discharge can be modified by varying the position of the plasma torch from a fixed substrate.

## 2. Experimental setup

### 2. 1. Characteristics of the plasma source

The post-discharge is generated by an RF atmospheric plasma torch from SurfX Technologies (Atomflo$^{TM}$ 400L-Series) [Babayan-2008]. The controller of the plasma source includes an RF generator (27.12 MHz), an auto-tuning matching network and a gas delivery system with two mass-flow controllers to regulate the helium and oxygen gases fuelling the plasma torch. The flow rate of helium (carrier gas) and oxygen (reactive gas) can be adjusted from 10 to 20 L/min and from 0 to 0.8 L/min respectively. As presented in [Figure 1.a.], the resulting gas mixture enters through a tube attached to a rectangular housing. Inside, two perforated sheets uniformize the gas flow down the housing. Then, the gas flows around the left and right edges of the upper electrode and passes through a slit in the center of the lower electrode. Plasma is struck and maintained between these electrodes by applying an RF power to the upper electrode while the lower electrode is grounded. The RF power commonly used is comprised between 60W and 160 W. The geometry of the slit is described as "linear" due to the ratio of its aperture length (20 mm) to its width (0.8 mm). For all the experiments performed in this paper, the plasma torch was always turned on 20 minutes before any measurement to ensure a steady-state regime and so no thermal effects influences issued from the progressive heating of the plasma torch.

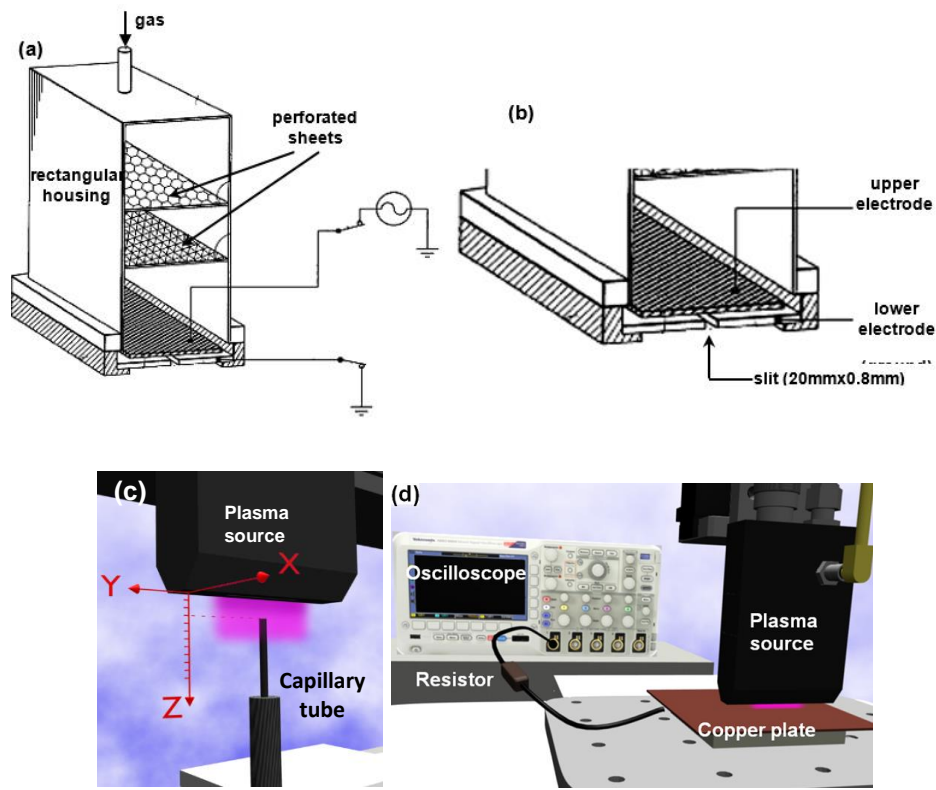

Figure 1. (a) Cross sectional diagram of the RF plasma source (Patent US7329608 Surfx Technologies [Babayan-2008]) (b) Zoom of a (c)Disposition of the capillary tube to the plasma torch for the acquisition of mass spectra (d) Experimental set up to measure DC currents in the flowing post-discharge.

## 2.2. Diagnostics

Mass spectra of the post-discharge were acquired by a quadripolar mass spectrometer (Balzers QMS 200) coupled with a turbomolecular pump (Pfeiffer TSU 062H) to a base pressure of approximately $1.10^{-8}$ mbar. The mass spectrometer measured the gaseous species in the post-discharge through a capillary tube, placed in parallel with the gas flow, as illustrated in [Figure 1.c.]. The capillary tube is 1m in length, with an inner diameter of 0.01 cm. It is heated by a resistance to maintain the gases temperature at 150°C, thus avoiding their condensation before entering in the ionization chamber. The ionization energy was set to 70 eV, as it is usually found in the literature [Glosik-1978], [Stoffels-2006].

Optical emission spectroscopy was performed with the SpectraPro-2500i spectrometer from ACTON research Corporation (0.500 meter focal length, triple grating imaging). Light emitted by the post-discharge was collected by an optical fiber and transmitted to the entrance slit (50 µm) of the monochromator. There, the light is collimated, diffracted, focused on the exit slit and finally captured by a CCD camera from Princeton Instruments. Each optical emission spectrum was acquired with the 1800 grooves.mm$^{-1}$ grating (blazed at 500 nm) and recorded on 30 accumulations with an exposure time of 25 ms. Experimentally, we observed that an increase in the $O_2$ flow rate was always inducing a decrease in all the emission lines/bands of the post-discharge. As a consequence, from an oxygen flow rate to another one, the deexcitation of a specific species was always proportional to the overall decay of the post-discharge emission. To solve this problem, for every $O_2$ flow rate, the emissions of all the species were divided by the emission of the whole post-discharge (i.e. a continuum ranging from 250 nm to 850 nm).

Line-emission absorption spectroscopy was applied to evidence He ($2^3S$) metastable states, more precisely on the transition $2^3S$-$3^3P$ at 388.9nm. This method was reliable and only required an external light source emitting the appropriate wavelength. The absorption rate A is related to the emission intensities by the relation $A=(I_L+I_P-I_{L+P})/I_L$, where $I_P$ is the light intensity emitted from the post-discharge, $I_L$ is the line intensity from the external lamp and $I_{L+P}$ is the line intensity from the two light sources [Gavare-2006], [Li-2005]. The light intensities were measured by placing the optical fiber along the Y-axis (see [Figure 1.c.]), 1cm away from the post-discharge.

To measure a DC current in the post-discharge, we introduced a copper plate downstream, as illustrated in [Figure 1.d.]. We called "gap" the distance separating the plasma source's top head from the copper plate. The current was obtained by measuring the potential difference across a resistor (820 kΩ) connecting the copper plate to an oscilloscope. For an accurate representation of the signals, we used a digital phosphor oscilloscope from Tektronix (DPO 3032) with a band-gap of 300 MHz and a sample rate as high as 2,5 GS/s.

## 3. Results & Discussion

In this section, we discuss how the current measured in the flowing post-discharge is depending on the $O_2$ flow rate, the helium flow rate and on the gap (distance separating the plasma source from a substrate placed downstream). The measurement of a DC current in the post-discharge is correlated with chemistry reactions evidenced by means of mass spectrometry, optical emission/absorption spectroscopy.

## 3. 1. Influence of the $O_2$ flow rate – Results from MS

**3.1.1. Results from MS**

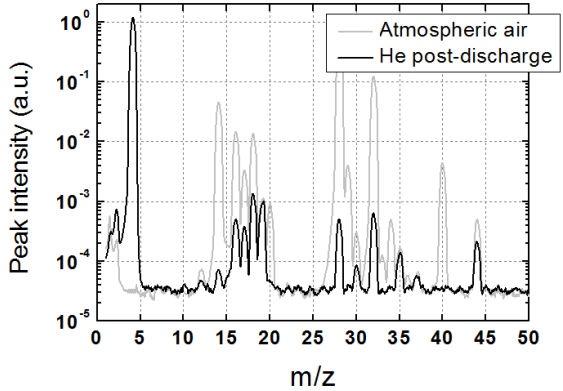

*Figure 2. Mass spectrum of the atmospheric air (grey curve) and mass spectrum of the flowing post-discharge (black curve) only supplied in helium gas, with $\Phi(He)$=15 L/min, $P_{RF}$=120W and $Z_{capillary}$=1 mm.*

The mass spectra of the air (grey curve) and of the flowing post-discharge supplied in helium gas (black curve) are represented in [Figure 2]. The mass spectrum of the air shows m/z ratios at 14, 16, 28 and 32 corresponding to the N, O, $N_2$ and $O_2$ species respectively. $N_2$ and $O_2$ are basic compounds of the air which is not the case for atomic O and N. Actually, the peaks at 14 and 16 represent $N^+$ and $O^+$ species (and/or doubly charged molecular ions $N_2^{2+}$ and $O_2^{2+}$) produced in the ionization chamber of the mass spectrometer. The high electron energy (70 eV) allows the fragmentation of the $O_2$ and $N_2$ molecules by dissociative ionization processes. Peaks measured at 14 and 16 represent a background (intrinsic to the ionization chamber). The same remark applies for the peak at m/z=30 indicating the production of nitric oxide molecules (NO) in the ionization chamber as a product of atomic O and N. The mass spectrum of the post-discharge (black curve in Figure 2]) is obtained for a distance of 1 mm separating the capillary tube from the plasma torch. This mass spectrum shows that the amounts of $N_2$ and $O_2$ become lower and the products of their reactions (m/z=14,16,30) as well. Moreover the argon peak (m/z=40) disappears while the $CO_2$ peak (m/z=44) is reduced in size. Of course, a very elevated peak of helium is measured at m/z=4.

Now that we have evidenced the fragmentation phenomenon of molecules in the ionization chamber, a flowing post-discharge only supplied with $\Phi(He)$=15 L/min ([Figure 3.a.]) is compared to a flowing post-discharge supplied with $\Phi(He)$=15 L/min and $\Phi(O_2)$=300 mL/min ([Figure 3.b.]). For these two cases, the peaks of helium, nitrogen and oxygenated species have been spatially measured by positioning the capillary tube at different positions along the Z-axis of the plasma source, as indicated in [Figure 1.c.].

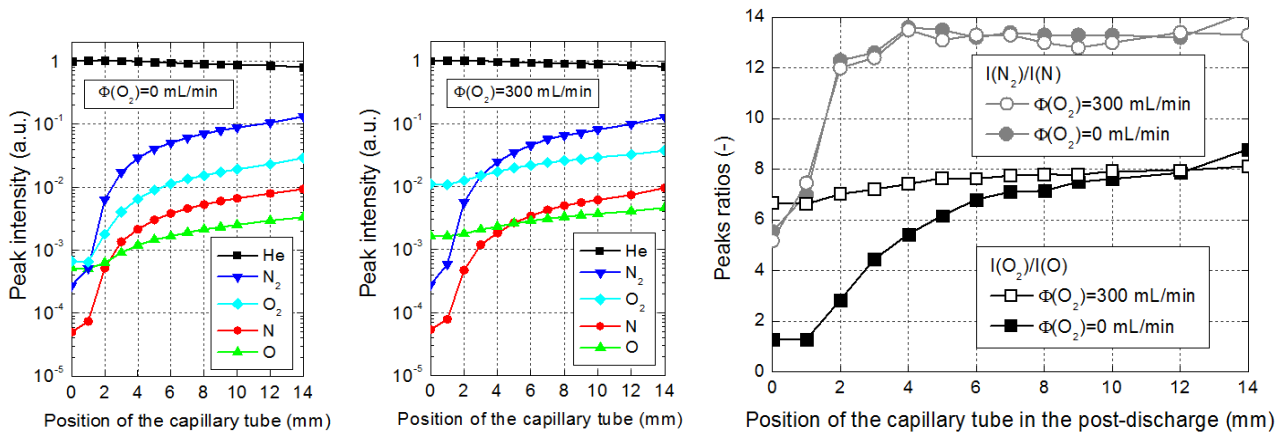

Figure 3. Spatial resolution along the Z-axis of the He, N, $N_2$, O and $O_2$ peaks intensities for $\Phi(He)$=15 L/min, $P_{RF}$=120W and (a) $\Phi(O_2)$=0 mL/min (b) $\Phi(O_2)$=300 mL/min. (c) Peaks ratios of nitrogen species and oxygen species.

In [Figure 3.a.], the chemical composition of the post-discharge is not homogeneous along the Z-axis. For a position of the capillary tube higher than Z=2 mm, the nitrogen and oxygenated species are no more negligible and represent potential sources of contamination in the treatment of surfaces. In [Figure 3.b.], the injection of $O_2$ at 300 mL/min causes the production of $O_2$ and O in the post-discharge. The comparison between [Figure 3.a.] and [Figure 3.b.] clearly shows two distinct atomic oxygen profiles along the Z-axis. Two sources of atomic oxygen must be considered: the fragmentation of the atmospheric $O_2$ in the ionization chamber (which applies in both cases) and the plasma source (which only applies in the second case). However, the two atomic nitrogen profiles show no difference thus indicating that they are only produced in the ionization chamber of the mass spectrometer. In [Figure 3.c.], the $I(O_2)/I(O)$ ratios for $\Phi(O_2)$=0 mL/min and for $\Phi(O_2)$=300 mL/min have been plotted for different positions of the capillary tube in the flowing post-discharge. The two curves are significantly different, thus evidencing the dissociation of $O_2$ by the plasma source. The same peaks ratios have been plotted for the nitrogen species ($I(N_2)/I(N)$). With/without a $O_2$ flow rate, the two curves still match, thus attesting that $N_2$ is not dissociated into atomic nitrogen species in the post-discharge.

As those nitrogen and oxygenated species exist in the flowing post-discharge, they represent a potential source of contamination in the treatment of surfaces. We compared this contamination to the contamination measured in a usual low-pressure plasma chamber. As illustrated in [Figure 4.], the peaks intensities of the nitrogen and oxygenated species have been plotted versus the pressure in the case of the RF plasma torch (open symbols) and in the case of a pyrex bell jar covering a stainless base connected to a primary pumping (filled symbols). The pressure in the vacuum chamber was controlled by a Baratron gauge for a set of pressures ranging from 1 Torr to 760 Torr (atmospheric pressure). This figure shows that the contamination in our atmospheric post-discharge (for Z<2 mm) is similar to the contamination in a vacuum chamber working under a pressure of about 1 Torr. In other words, the surface contamination induced by the flowing post-discharge during the process is controlled over time, potentially negligible on the treatment of surfaces, but not necessarily negligible on the plasma chemistry as we will discuss now through OES and OAS experiments.

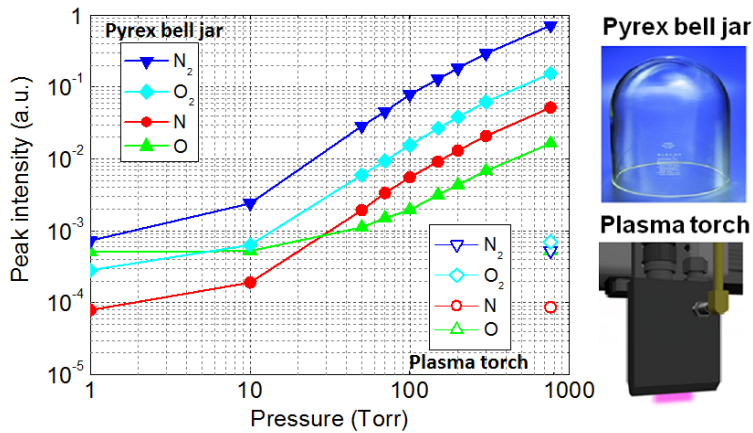

*Figure 4. Peaks intensities of nitrogen and oxygenated species versus the pressure for two cases: the RF plasma torch (operating only at 760 Torr for Φ(He)=15 L/min, Φ(O$_2$)=0 mL/min, P$_{RF}$=120W and Z$_{capillary}$=1 mm) and a usual vacuum chamber.*

## 3.2. Influence of the O$_2$ flow rate – Results from OES and OAS

In [Figure 5ab], the optical emissions of several species are plotted versus the O$_2$ flow rate: N$_2$ (337 nm), $N_2^+$ (391 nm), He (3$^3$S, 706.5 nm), O (3$^3$S, 844.6 nm), OH (310 nm), $O_2(b^1\Sigma_g^+, 765\,nm)$ and $O_2^+$ (525 nm). Production of O, $O_2^+$ and $O_2(b^1\Sigma_g^+)$ species (increasing curves) is balanced by the consumption of He, OH, N$_2$ and $N_2^+$ (decreasing curves). We have also reported in [Figure 6] the emission spectra of $O_2(b^1\Sigma_g^+)$, $N_2^+$ and $O_2^+$ as they are rarely presented in the literature when measured in a post-discharge. The band of the singlet sigma metastable oxygen $O_2(b^1\Sigma_g^+)$ was measured between 758 nm and 770 nm; it lies closely above the $O_2(a^1\Delta_g)$ excited singlet state and $O_2(X^3\Sigma_g^-)$ triplet ground state [Jeong-1998], [Minaev-1997], [Schmidt-1999]. To evidence the 2$^3$S metastable states of helium, we have reported in [Figure 7] its absorption rate versus the O$_2$ flow rate obtained by line-emission absorption spectroscopy on the transition He(3$^3$P-2$^3$S) at 388.9nm. All those results will be now discussed for each species to draw up the main chemistry mechanisms in the flowing post-discharge.

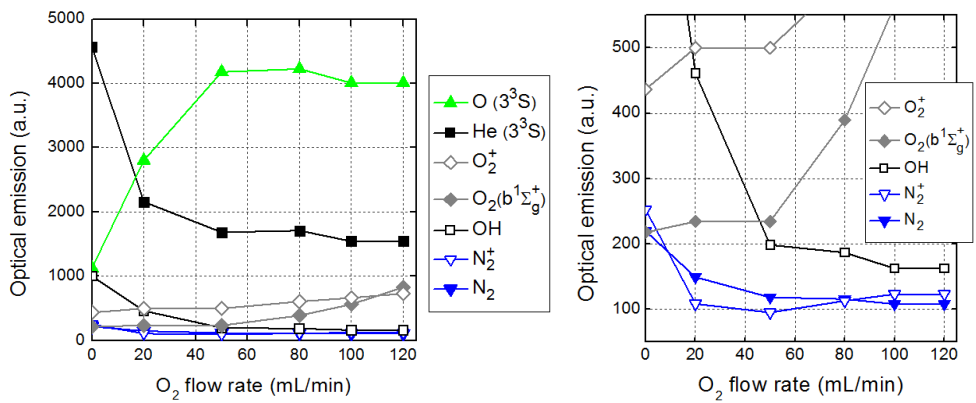

*Figure 5. (a) Optical emission spectra of the whole post-discharge versus the O$_2$ flow rate for Φ(He)=15 L/min, P$_{RF}$=120W and gap=5 mm. (b) Zoom of figure 5.a.*

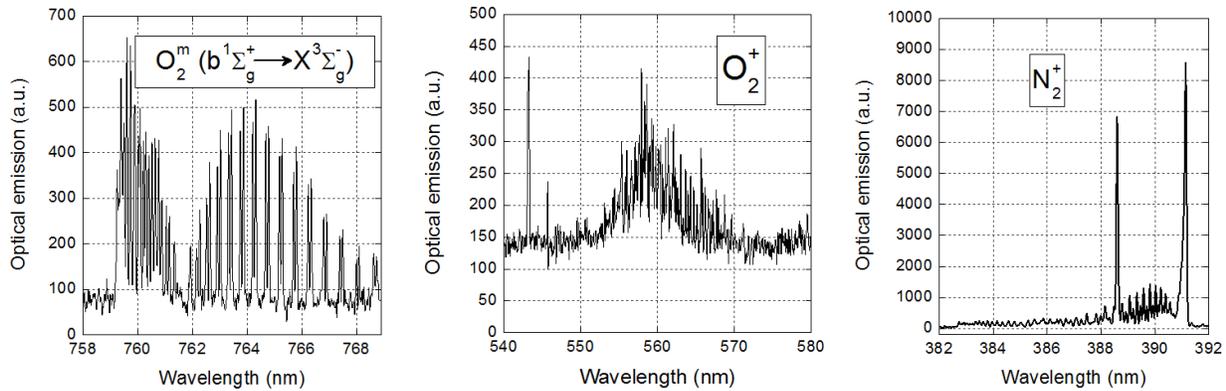

Figure 6. Optical emission spectra of $O_2$ metastables, $O_2^+$ ions and $N_2^+$ ions measured in the flowing post-discharge for the following experimental conditions (a & b) $\Phi(He)$=15 L/min, $\Phi(O_2)$=100 mL/min, $P_{RF}$=120W, gap=6 mm, (c) $\Phi(He)$=15 L/min, $\Phi(O_2)$=0 mL/min, $P_{RF}$=120W, gap=5 mm.

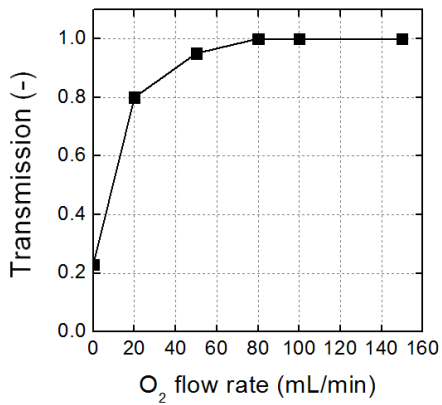

Figure 7. Transmission of He $2^3S$ at 388.9 nm, in the He-$O_2$ flowing post-discharge with $\Phi(He)$=15 L/min, $\Phi(O_2)$=0 & 100 mL/min and $P_{RF}$=120 W.

*$O_2^+$ ions*

From 0 to 120 mL/min, the increase in the $O_2$ flow rate is accompanied by a permanent increase in $O_2^+$ ions, as illustrated in [Figure 5.b.]. Several mechanisms are usually considered to explain the production of $O_2^+$ but all cannot apply here. Thus, a charge transfer reaction involving $O_2$ molecules and $O^+$ ions is not possible since no emission of $O^+$ was detected at 465 nm [Léveillé-2006]. A dissociative charge transfer with $He_2^+$ ions is not possible either since no optical emission was detected from the $He_2$ band (3p$\pi$e$^3\Pi_g$-2s$\sigma$a$^3\Sigma_u^+$) between 462 nm and 468 nm [Kutasi-2001]. The electron impact ionization of $O_2$ (Reaction [A] in Table 1) is possible but seems negligible due to the low value of its rate constant and also because of the usual weak electron densities measured in similar He-$O_2$ RF post-discharges (between $8.10^{10}$ cm$^{-3}$ and $1.3.10^{12}$ cm$^{-3}$ [Gonzalez-2010]). The main mechanism explaining the production of $O_2^+$ ions is the Penning ionization of $O_2$ molecules by He metastables (Reaction [B] in Table 1). Such a reaction requires He ($2^3S$) metastables that will be evidenced afterwards, and produces – in addition to $O_2^+$ ions – electrons and radiative states of helium.

*$N_2^+$ ions*

At atmospheric pressure, the production of $N_2^+$ ions could be attributed to three known mechanisms: a charge transfer from $He_2^+$ ions to $N_2$, a charge transfer from $He^+$ ions to $N_2$ and the Penning ionization of $N_2$ from metastable helium (Reaction [C]). As we did not detect any optical emission from $He_2$ and $He^+$ species in our experiments, the reaction [C] is considered as the main source of $N_2^+(B^2\Sigma_u^+)$ ions. This last reaction has been widely studied in low-pressure experiments [Jolly-1980], [Cher-1969], [Taieb-1976], [Collins-1986] and is considered at atmospheric pressure as the major mechanism for the production of $N_2^+(B^2\Sigma_u^+)$ ions [Chan-2011]. Parallel to this emission at 391 nm, we observed the emission of $N_2^+(X^2\Sigma_g^+)$ species on the (0-0) band of the first negative system at 391.4 nm (Reaction [D]). The [Figure 5.b.] also indicates in which extent the $O_2$ species influence the Penning ionization of $N_2(X^1\Sigma_g^+)$. Thus, a flow as low as 20 mL/min in $O_2$ reduces by more than 50% the emission of $N_2^+(B^2\Sigma_u^+)$. The explanation to this decrease can be found in the absorption peak of He ($2^3$S) represented in [Figure 7]. The absorption rate is drastically reduced by increasing the $O_2$ flow rate. He ($2^3$S) metastables are consumed as mentioned in the reaction [B], thus becoming a limiting reactant in reaction [C]. As the reaction [C] becomes less preponderant, the emission of $N_2^+(B^2\Sigma_u^+)$ decreases with the increase in the $O_2$ flow rate.

*Atomic oxygen*

For $\Phi(O_2)$=0-50 mL/min, the increase in the O ($3^3$S) emission is consistent with our previous results from mass spectrometry ([Figure 3 (a&b)]) in which atomic oxygen was only produced when $O_2$ was injected in the discharge. The direct dissociation of $O_2$ molecules through collisions with energetic electrons is obviously inefficient, because of the electronegative nature of $O_2$ [Khan-2008]. The most probable channel to produce atomic oxygen is the Penning ionization of $O_2$ molecules (Reaction [B]) followed by the electron impact dissociation of $O_2^+$ (Reaction [E]). The rate constant of this last reaction ($4.8.10^{-7}$ cm$^3$.mol$^{-1}$.s$^{-1}$) is particularly elevated. For $\Phi(O_2)$=50-120 mL/min, the optical emission of O ($3^3$S) has reached a plateau that can be interpreted as resulting from an equilibrium between production and consumption mechanisms.

*Helium species*

The Grotrian diagrams of He I and O I indicate that radiative states of helium present energetic levels almost twice higher than those of oxygen [Moore-1968]. It is suspected that the addition of 1% of an electronegative gas, namely $O_2$, in the inert helium gas causes a reduction of the electron density so that the electron collisionnal processes can no more participate to the excitation of the He radiative states [Léveillé-2005]. In [Figure 5.a.], for $O_2$ flow rates increasing from 0 mL/min to 50 mL/min, we have registered a decay in the emission of He ($3^3$S) counterbalanced by the production of radiative O ($3^3$S) at 844.6 nm. The same emission decay from He radiative states was measured by increasing the $O_2$ flow rate of an atmospheric plasma jet sustained by helium [Léveillé-2006]. Moreover, another experiment performed in a He-$O_2$ microwave discharge at atmospheric pressure, indicates a decrease in the density of He ($2^3$S) metastable states as a function of the oxygen concentration in helium, thus following – according to the authors – a similar optical emission trend as the excited He atoms at 706 nm [Cardoso-2006]. Now, for $O_2$ flow rates increasing from 50 mL/min to 120 mL/min, the optical emission of the He ($3^3$S) remains almost constant which is also consistent with the plateau of the O ($3^3$S) species.

*$O_2$ mestastable molecules*

For $\Phi(O_2)$= 0-50 mL/min, the reactions [B] and [F] apply so that all the injected $O_2$ molecules are both Penning-ionized to produce $O_2^+$ ions and excited by electrons to produce $O_2(b^1\Sigma_g^+)$ species. According to their respective rate constants, the production of $O_2^+$ ions is more efficient than the production of $O_2(b^1\Sigma_g^+)$ metastables, which is consistent with their optical emissions: for $\Phi(O_2)$ increasing from 0 to 50 mL/min, the emission of $O_2^+$ is always higher than the emission of $O_2(b^1\Sigma_g^+)$. Beyond 50 mL/min, the drastic increase in the emission of $O_2(b^1\Sigma_g^+)$ let us think that another channel for the production of $O_2(b^1\Sigma_g^+)$ is opened. The $O_2$ molecules that were all consumed by the reactions [B] and [F] are from now on in excess, colliding with O ($^1$D) atoms to produce atomic oxygen and $O_2(b^1\Sigma_g^+)$ species, as reported in the reaction [G]. $O_2(a^1\Delta_g)$, which is another metastable state of oxygen, is known to have a lifetime much more longer than $O_2(b^1\Sigma_g^+)$ [Popovic-2010]. The corresponding transition $2a^1\Delta_g \rightarrow 2X^3\Sigma_g^-$ is commonly measured at 1.27 µm by infrared emission spectroscopy or in the visible range at 634 nm and 703 nm by OES [Sousa-2011]. Despite its very long lifetime, no optical emission of $O_2(a^1\Delta_g)$ was detected in the flowing post-discharge, neither at 634 nm, nor at 703 nm. The absence of this metastable radiative decay can be explained by the existence of other processes which consume more efficiently the $O_2(a^1\Delta_g)$: either by quenching with $O_2$ molecules (Reaction [H]), and/or to a lesser extent by electronic dissociative attachment (Reaction [I])

|   | Reaction | Rate constant (cm$^3$.mol$^{-1}$.s$^{-1}$) | Ref. |
|---|---|---|---|
| A | $O_2 + e \rightarrow 2e + O_2^+$ | 3.3.10$^{-15}$ | [Lee-1994] |
| B | $He(2^3S) + O_2 \rightarrow He + O_2^+ + e$ | 2.4.10$^{-10}$ | [Lee-2005] |
| C | $He(2^3S) + N_2(X^1\Sigma_g^+, v=0) \rightarrow He + N_2^+(B^2\Sigma_u^+, v'=0) + e$ | 6.9.10$^{-11}$ | [Léveillé-2006] |
| D | $N_2^+(B^2\Sigma_u^+, v'=0) \rightarrow N_2^+(X^2\Sigma_g^+, v=0) + h\nu$ | - | [Naveed-2006] |
| E | $O_2^+ + e \rightarrow 2O$ | 4.8.10$^{-7}$ | [Lee-2005] |
| F | $O_2 + e \rightarrow O_2(b^1\Sigma_g^+) + e$ | 3.1.10$^{-26}$ | [Gudmundsson-2004] |
| G | $O(^1D) + O_2 \rightarrow O + O_2(b^1\Sigma_g^+)$ | 3.08.10$^{-11}$ | [Ionin-2007] |
| H | $O_2(a^1\Delta_g) + O_2 \rightarrow O_3 + O$ | 2.9.10$^{-21}$ | [Datta-1979] |
| I | $O_2(a^1\Delta_g) + e \rightarrow O^- + O$ | 2.3.10$^{-22}$ | [Hicks-2005] |
| J | $O_2 + O^+ \rightarrow O_2^+ + O$ | 2.0.10$^{-10}$ | [Lee-1994] |
| K | $N_2 + O \rightarrow NO + N$ | 3.9.10$^{-22}$ | [Léveillé-2005] |

*Table 1. Partial list of reactions with their rate constants for the He-O$_2$ flowing post-discharge.*

<u>*OH molecules*</u>

At atmospheric pressure, hydroxyl radicals (OH) are mainly produced by electron-impact dissociation of H$_2$O molecules (single step process) or by electron-impact ionization of H$_2$O followed by dissociation of H$_2$O$^+$ to produce OH (two-step process) [Goree-2006], [Ono-2002], [Khacef-2002]. As those reactions require energetic electrons (higher than 2 eV), they probably do not occur in the flowing post-discharge [Itikawa-2005], [Gonzales-2010]. The production of the OH radicals is assumed to be performed between the electrodes, where H$_2$O molecules are adsorbed when the plasma torch is not operating. Moreover, the injection of O$_2$ in the post-discharge is assumed to decrease the electron density, thus limiting the production of hydroxyl radicals and therefore their optical emission. For this reason, a decrease in the OH emission is observed in [Figure 5.a.] for increasing O$_2$ flow rates.

*Species that have not been observed*

In agreement with our previous results from mass spectrometry, no optical emission from atomic nitrogen could be observed at 575 nm or 746 nm. No emission from $O^+$ ions could be observed a 465 nm because either they do not exist, or they are quenched by $O_2$ molecules according to the reaction [J]. The optical emissions from the Rydberg-Rydberg transitions of the NO molecule have not been observed either [Rosen-1970], since the only mechanism allowing its formation is the reaction [K] requiring a gas temperature higher than 1400 K. This temperature is quite higher than the one we measured in the post-discharge (almost 400 K), obtained by fitting the experimental OH band (310 nm) to the OH model computed on the LIFBASE software. The study of the ozone ($O_3$) concentration by optical absorption spectroscopy has not been carried out due to the small size of the post-discharge. It could participate to the production of other species but it is not directly suspected to sustain the DC current measured in the flowing post-discharge.

Those OES results have highlighted the existence of excited and positive charged species in the flowing post-discharge and suggested mechanisms explaining their production and consumption rates. Two separate Penning ionizations have been highlighted as the major mechanisms allowing the productions of $N_2^+$ and $O_2^+$ species. According to their rate coefficients, the Penning ionization induced by $O_2$ (k=2.4.10$^{-10}$ cm$^3$.mol$^{-1}$.s$^{-1}$) is faster than the Penning ionization induced by $N_2$ (k=6.9.10$^{-11}$ cm$^3$.mol$^{-1}$.s$^{-1}$). For this reason, the Penning ionization with $O_2$ molecules quenches the majority of He metastable atoms. As a consequence, very few of them remain available for the Penning ionization of the $N_2$ molecules, thus explaining why in [Figure 5.a.] the optical emission of $N_2^+$ is lower than the optical emission of $O_2^+$.

## 3.3. Influence of the $O_2$ flow rate – Electrical measurements

**3.3.1. Evidence of a DC current measured in the post-discharge**

We have evidenced the production of positive charged species ($O_2^+$, $N_2^+$) which, combined to the gas flow sustained by the plasma torch, could generate an electrical current. To evidence this current, a copper plate was placed 1 mm downstream from the plasma torch to collect the charged species carried away by the flowing post-discharge (see [Figure 1.d.]). Then, the resulting current was measured across a resistor connecting the copper plate to the oscilloscope.

The variation of the post-discharge current versus time is plotted in [Figure 8.] for Φ(He)=15 L/min and an RF power of 120W. The main feature of this current is the absence of any RF frequency or even AC component. All are pure DC currents, induced by a permanent transport of charged species within the flowing post-discharge. The instant t=0 min corresponds to the injection of a $O_2$ flow rate of 100 mL/min. At t=0 min, the current measured is 4 µA and increases drastically to 21 µA in less than 1 second. Then, until t=4 min it decreases following a first-order exponential decay to finally reach the value of 12 µA. This decay is very long (several minutes) and corresponds to a transition regime not very well understood yet and beyond the scope of this paper. For this reason, all the currents plotted afterwards in this paper were only measured at least 5 minutes after the beginning of the injection of $O_2$ to be sure that a permanent regime was reached and the measurements reproducible.

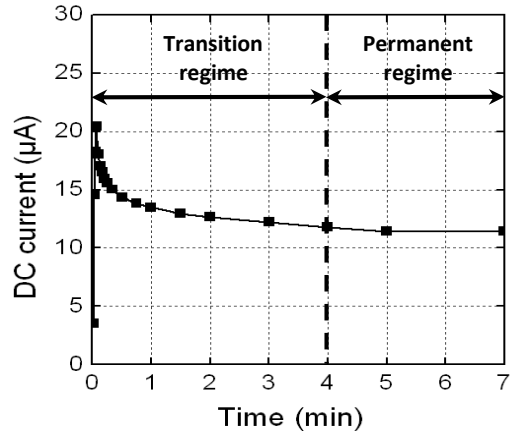

*Figure 8. Evolution of the post-discharge DC current versus time, for $\Phi(He)$=15 L/min, $\Phi(O_2)$=100 mL/min, $P_{RF}$=120W and gap =1 mm. The $O_2$ gas was injected at t= 0 min.*

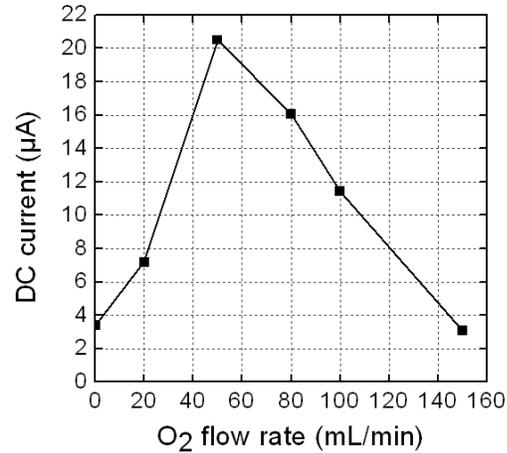

*Figure 9. Variations of the DC current measured in the steady flowing post-discharge, versus the $O_2$ flow rate for $\Phi(He)$=15 L/min, $P_{RF}$=120W and gap =1 mm.*

### 3.3.2. Experimental study of the permanent regime

The effect of the $O_2$ gas on the DC current has been investigated for flow rates ranging from 0 to 150 mL/min with a helium flow rate fixed at 15 L/min and an RF power at 120W. The [Figure 9.] shows two operating regimes: for $O_2$ flow rates increasing from 0 to 50 mL/min, the DC current increases from 3 to 20.5 µA and for $O_2$ flow rates increasing from 50 to 150 mL/min, it decreases from 20.5 to 3µA. During the first operating regime, the Penning ionization induced by $O_2$ (Reaction [B]) becomes always more efficient until a critical flow rate fixed at $\Phi(O_2)$=50 mL/min and for which a maximum current is reached (20.5 µA). Just for recall, this critical flow rate was previously mentioned concerning the OES results in [Figure 5ab.], especially because it was marking the end of the $O_2$ metastables plateau and the beginning of the atomic oxygen plateau.

Beyond this threshold (50 mL/min), the current decreases while according to the optical emissions in [Figure 5.b.]: (i) the production of $O_2^+$ ions increases and (ii) the production of atomic oxygen is leveled on a plateau. A mechanism that could explain the decrease in the current is a rate competition between the permanent production of $O_2^+$ ions and the production of negative charged species (such as $O^-$, $O_2^-$). The production of oxygen negative ions seems a consistent assumption because species with high electronegativity such as O tend to form electronegative ions by the recombination of electrons and radicals at low energies [Yi-2003]. Moreover, charge neutralization between $N_2^+$, $O_2^+$ and oxygen negative ions may be elevated due to the atmospheric pressure. By increasing the $O_2$ flow rate, the charge neutralization in the plasma could become more efficient, thus decreasing the plasma density and therefore reducing the post-discharge current. The current measured on the copper plate would be the sum of a positive and a negative components counteracting once the threshold of 50 mL/min is reached.

Another reason that could explain the decrease in the DC current would be the observed reduction in size of the flowing post-discharge when the $O_2$ flow rate is increased. As the gap remains constant, reducing in size the post-discharge means a less efficient collection of charged species on the copper plate.

### 3.4. Influence of the He flow rate – Results from OES

The influence of the helium flow rate on the properties of the flowing post-discharge have been investigated by OES. In [Figure 10.], the optical emissions of the previous species are plotted versus the helium flow rate for values ranging from 10 to 20 L/min. The $O_2$ flow rate is fixed at 100 mL/min and the RF power still at 120 W. The only increasing curve corresponds to the emission of the atomic oxygen at 777 nm. It represents a more significant convective transport of the atomic oxygen along the flow axis of the post-discharge. This phenomenon has already been observed in the case of a He-$O_2$ plasma-jet [Léveillé-2006] and can be explained by the difference of energy between the upper energy level of the O (777 nm) and the upper energy level of He (706 nm): 10.74 eV and 22.72 eV, respectively.

All the other optical emissions slightly decrease versus $\Phi(He)$. The slight decrease in the emission of $N_2^+$ ions is attributed to the fact that by increasing the helium flow rate, the post-discharge grows bigger and the nitrogen contamination lower. As a consequence, the $N_2(X^1\Sigma_g^+)$ species involved in the Penning ionization (Reaction [C]) limit this reaction, and a slight decrease in the emission of the $N_2^+$ ions is then observed. The slight decay in the emission of the $O_2^+$ ions operates on the same principle: the growing post-discharge slightly reduces the atmospheric $O_2$ component that could participate in the Penning ionization (Reaction [B]).

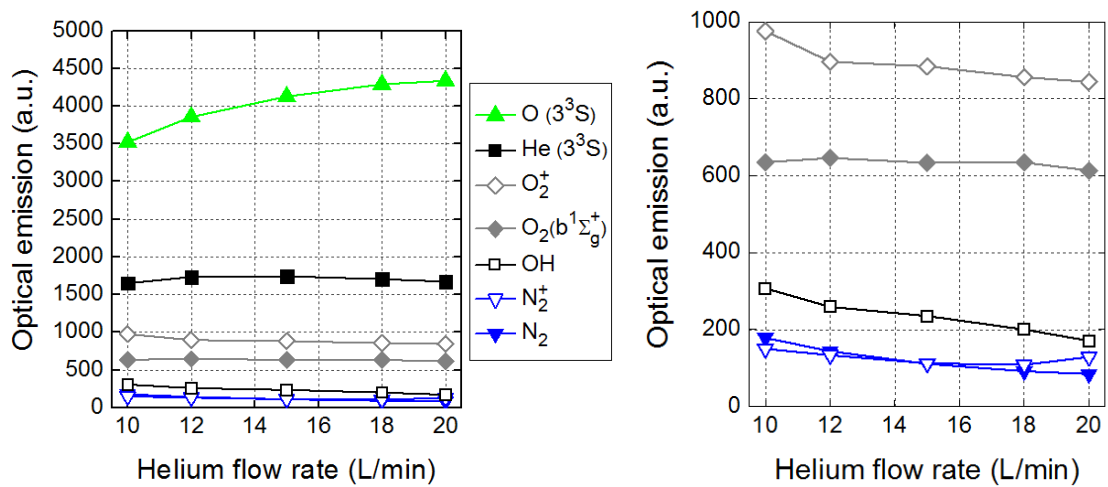

Figure 10. (a) Optical emissions of O, He, $O_2^+$, $O_2^*$, OH, $N_2$ and $N_2^+$ versus the helium flow rate for $\Phi(O_2)$=100 mL/min, $P_{RF}$=120 W and gap = 5 mm. (b) Zoom of "(a)".

### 3.5. Influence of the He flow rate – Electrical measurements

In [Figure 11], for a $O_2$ flow rate as high as 100 mL/min injected in the flowing post-discharge, the DC current measured for $\Phi(He)$=10-20 L/min, slightly decreases from 19 µA to 16.5 µA. This behavior is consistent with the optical emissions of the $N_2^+$ and $O_2^+$ ions, also slightly decreasing in [Figure 10.]. The same behavior is also obtained without supplying the post-discharge in oxygen gas; the resulting DC current is again slightly decreasing but for values approximately 4 times lower than in the previous case.

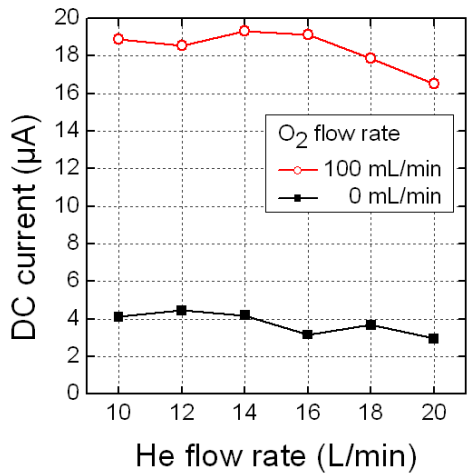

Figure 11. Variations of the DC current in the steady flowing post-discharge versus the helium flow rate for $P_{RF}$=120W and gap= 1mm.

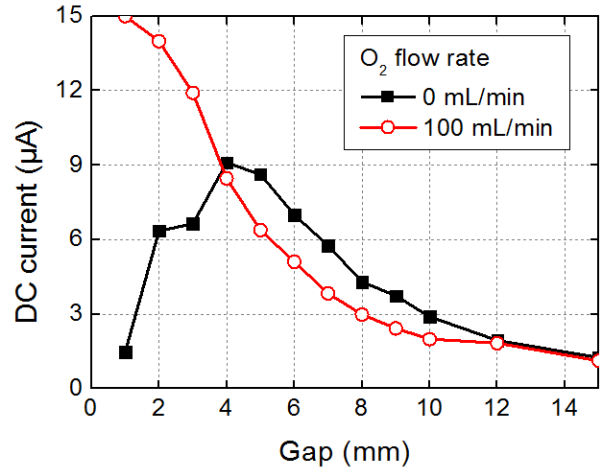

Figure 12. Variations of the DC current versus the gap (distance torch-surface) for $\Phi(He)$=15 L/min, $P_{RF}$=120 W and $\Phi(O_2)$=0-100 mL/min.

### 3.6. Influence of the torch/substrate distance: electrical measurements

The interaction of the flowing post-discharge with a surface (the copper plate) has been investigated through electrical measurements. As illustrated in [Figure 12], the DC current has been measured for different gaps ranging from 1 to 15 mm in two cases: with a $O_2$ flow rate of 100 mL/min (open symbols) and without injection of $O_2$ in the post-discharge (filled symbols).

For $\Phi(O_2)$=100 mL.min$^{-1}$, the DC current which is about 15 µA for a gap of 1mm drops to a value less than 1 µA for gaps higher than 14 mm. By increasing the gap, the copper plate collects less positive charged species ($N_2^+$ and $O_2^+$ ions), thus inducing an exponential decay of the DC current. It is also worth mentioning that whatever is the gap, the DC current remains always positive: the positive charged species have therefore a lifetime assumed to be longer than the one of the potential oxygenated negative charged species.

For $\Phi(O_2)$=0 mL.min$^{-1}$, the increase in the gap is not only reflected by a monotone decrease in the DC current. Such a decrease (filled symbols) is only confirmed for gaps comprised between 4 and 15 mm. On this range, this decrease overlays the previous current decay (open symbols) for the same reasons. However, for gaps comprised between 1 and 4 mm, the current clearly increases from 1 µA (at 1mm) to 9 µA (at 4 mm). Such an increase could be explained by two mechanisms: either a more efficient production of positive charged species or a more efficient consumption of oxygenated negative charged species. As no $O_2$ is injected in the plasma torch, this second assumption seems irrelevant. Therefore, for a helium flow rate fixed at 15 L/min, the increase in the gap from 1 to 4 mm is assumed to strengthen the turbulent regime of the flowing post-discharge. A better mixture with the atmospheric species is then obtained, in particular with the $N_2$ molecules, allowing subsequently a more efficient Penning ionization of $N_2$.

## 4. Conclusion

The flowing post-discharge generated by an RF He-$O_2$ plasma torch has been characterized by OES, MS and electrical measurements. We have evidenced that the contamination in the flowing post-discharge is similar to the contamination measured in a vacuum chamber at a pressure of 1 Torr. Even if increasing the helium flow rate reduces the atmospheric contamination on several decades, the atmospheric nitrogen and oxygenated species must still be taken into account on the chemistry processes in the flowing post-discharge.

By OES, we have identified the excited and positive charged species in the post-discharge and suggested mechanisms explaining the production of $N_2^+$ and $O_2^+$ ions. By electrical characterizations, we have evidenced the existence of a DC current depending mainly on the $O_2$ flow rate (and in a less extent to the helium flow rate). The current measured is considered as resulting from a competition between positive and negative charged species. The positive charged species mostly involved in the DC current are the $O_2^+$ ions. $N_2^+$ ions are also involved but in a less extent since the Penning ionization of $O_2^+$ is more efficient than the Penning ionization of $N_2^+$. The existence of negatively charged oxygenated species is assumed to play an important role for $O_2$ flow rates higher than 50 mL/min.

## 5. Acknowledgements

This work was part of the I.A.P (Interuniversitary Attraction Pole) program financially supported by the Belgian Federal Office for Science Policy (BELSPO). This work was also financially supported by the FNRS (Belgian National Fund for Scientific Research), Région Wallonne (OPTI2MAT Excellence Program) and the European Commission (FEDER – Revêtements Fonctionnels).